\documentclass{article}
\usepackage{mrs2005,epsfig}
\begin{document} 
\title{The Luminosity Function of Early-type Field Galaxies at $z\approx0.75$}

\author{
N.J.G. Cross$^{1,2}$ \& ACS Science Team$^{3}$.
}

\affil{
$^{1}$Institute for Astronomy, Royal Observatory,
Blackford Hill, Edinburgh, EH9 3HJ, UK. \\
$^{2}$Department of Physics and Astronomy, Johns Hopkins
University, 3400 North Charles Street, Baltimore, MD 21218, USA. \\
$^{3}$http://acs.pha.jhu.edu/general/personnel/sci-team/
}

\begin{abstract} 

We measure the luminosity function of morphologically selected E/S0
galaxies from $z=0.5$ to $z=1.0$ using deep high resolution Advanced 
Camera for Surveys (ACS) imaging data.  Our data extend 2 magnitudes 
deeper than the Deep Groth Strip Survey (DGSS). At $0.5<z<0.75$, we find 
$M_B^*-5\log\,h_{0.7}=-21.1\pm0.3$ and $\alpha=-0.53\pm0.2$, and at 
$0.75<z<1.0$, we find $M_B^*-5\log\,h_{0.7}=-21.4\pm0.2$. Our 
morphologically selected luminosity
functions are similar in both shape and number density to other
morphologically selected luminosity functions (e.g., DGSS), but we
find significant differences to the luminosity functions of samples 
selected using morphological proxies like colour or SED. The
difference is due to incompleteness from blue E/S0 galaxies, 
which make up to $\sim30\%$ of the sample and contamination from early-type
spirals. Most of the blue E/S0 galaxies have similar structural
properties to the red E/S0s and could passively evolve to form giant
red ellipticals at $z=0$. However, the bluest, $(U-V)_0<1.2$, have
much smaller Sersic parameters and would evolve into much fainter galaxies. 
These may be the progenitors of dwarf ellipticals. We demonstrate the
need for {\it both morphological and colour information} to constrain
the evolution of E/S0 galaxies.

\end{abstract} 

\vspace{-5mm} 
\section{Introduction} 

The luminosity function (LF) of galaxies is the number density of galaxies as
a function of absolute magnitude, and is often
parameterized using the Schechter Function \cite{Sch76}, where it is 
described by three numbers: $M^*$, the magnitude at which the 
number of bright galaxies rapidly decreases; $\phi^*$, the space density 
at $M^*$, and the faint end slope $\alpha$ which 
characterizes the ratio of dwarf galaxies to giant galaxies. Models of galaxy 
formation and evolution must be able to account for these parameters, which 
vary with galaxy type. Over the past few years, the luminosity function of 
high redshift ($z>0.5$) galaxies have been studied extensively through the 
use of deep, wide-area surveys.  Some of the more notable efforts include 
the Canada-France Redshift Survey (CFRS, \cite{Lll95}), the Calar Alto
Deep Imaging Survey (CADIS, \cite{Frd01}), the Deep Groth Strip Survey 
(DGSS, \cite{Im02}), and the Classifying Objects by Medium Band Observations 
(COMBO-17, \cite{Wlf03}, \cite{Bll04}). Most of these 
use deep, ground-based images with spectroscopic or photometric redshifts to 
construct the luminosity function, but do not have the spatial resolution to 
measure the structural properties of galaxies at higher redshifts. 

Without information on the structural properties, ground-based surveys have 
resorted to using colour information as a proxy for morphologies, whether this 
information comes in the form of a best-fit spectral energy distribution 
(e.g. \cite{Wlf03}, \cite{Bll04}), or a rest-frame colour cut 
(e.g. \cite{Lll95}). This can result in apparently discrepant results.  
For example, COMBO-17 originally found that the elliptical/S0 (E/S0) galaxies 
that produce $\sim50\%$ of the $z=0$ B-band luminosity density only 
contributed $\sim5\%$ at $z=1$. By contrast, using morphological
classification, the DGSS found that luminosity density
of E/S0s has increased by a factor of almost 2 over this range. Either
the luminosity of ellipticals has increased over time relative to
other types of galaxies or 
that the differences in colour-selection and morphological-selection have 
produced apparently inconsistent results between these surveys. 

Surveys using the Hubble Space Telescope (HST) such as the DGSS 
have been able to reliably morphologically classify 
and measure structural parameters for galaxies with $I_{AB}<22$ mag, but 
over much smaller areas of sky than the deep ground based surveys. 
These HST surveys have discovered a population of $0.3<z<1$ blue E/S0
galaxies (e.g. \cite{Mnt99}, \cite{Gbh03}) that have 
similar luminosities to standard red E/S0 galaxies, making up 
$30-50\%$ of the sample \cite{Mnt99}. Objects such as these 
demonstrate the inherent weakness of using  colour as a proxy for morphology. 
Dynamical masses suggest the these blue E/S0 galaxies are less 
massive than red E/S0 galaxies, \cite{Im01}, but resolution effects
make it hard to get reliable masses. Even so, the blue E/S0 galaxies 
may yet evolve into high mass red E/S0s through a combination of 
luminosity evolution that reddens the stellar population and dry
mergers that increase mass \cite{Fbr05}.

Luminosity evolution occurs when there is new star-formation, or when the 
stellar population ages, and does not necessarily imply any change in the
mass or number of stars in a galaxy. Colours can also be affected by
dust or AGNs. Structural parameters such as the size 
and shape are better indicators of the morphological evolution, since they 
are only weakly dependent on the age of the stellar population and are mainly
determined by dynamical characteristics such as total mass and angular 
momentum. The size and shape of the galaxy will not 
change significantly unless mass is added via mergers or
accretion. Small changes in the apparent shape and size
do occur when star-formation is localized, but these are much weaker 
changes than the variation in SED or colour. Thus morphology is a more
robust indicator of the nature of a galaxy: but it requires good
resolution. 

In this article, we examine the effect that colour and morphological
selection have on the measurement of the E/S0 luminosity function. We
use the improved resolution and sensitivity and field-of-view of the 
ACS, to determine the morphologies of
galaxies to deeper limits and over wider areas than before. We
use structural parameters to test whether blue E/S0 galaxies are 
progenitors of red E/S0 galaxies and what evolution has taken place 
from $z=1$ to $z=0.5$. 

\begin{figure}  
\vspace*{1.25cm}  
\begin{center}
\epsfig{figure=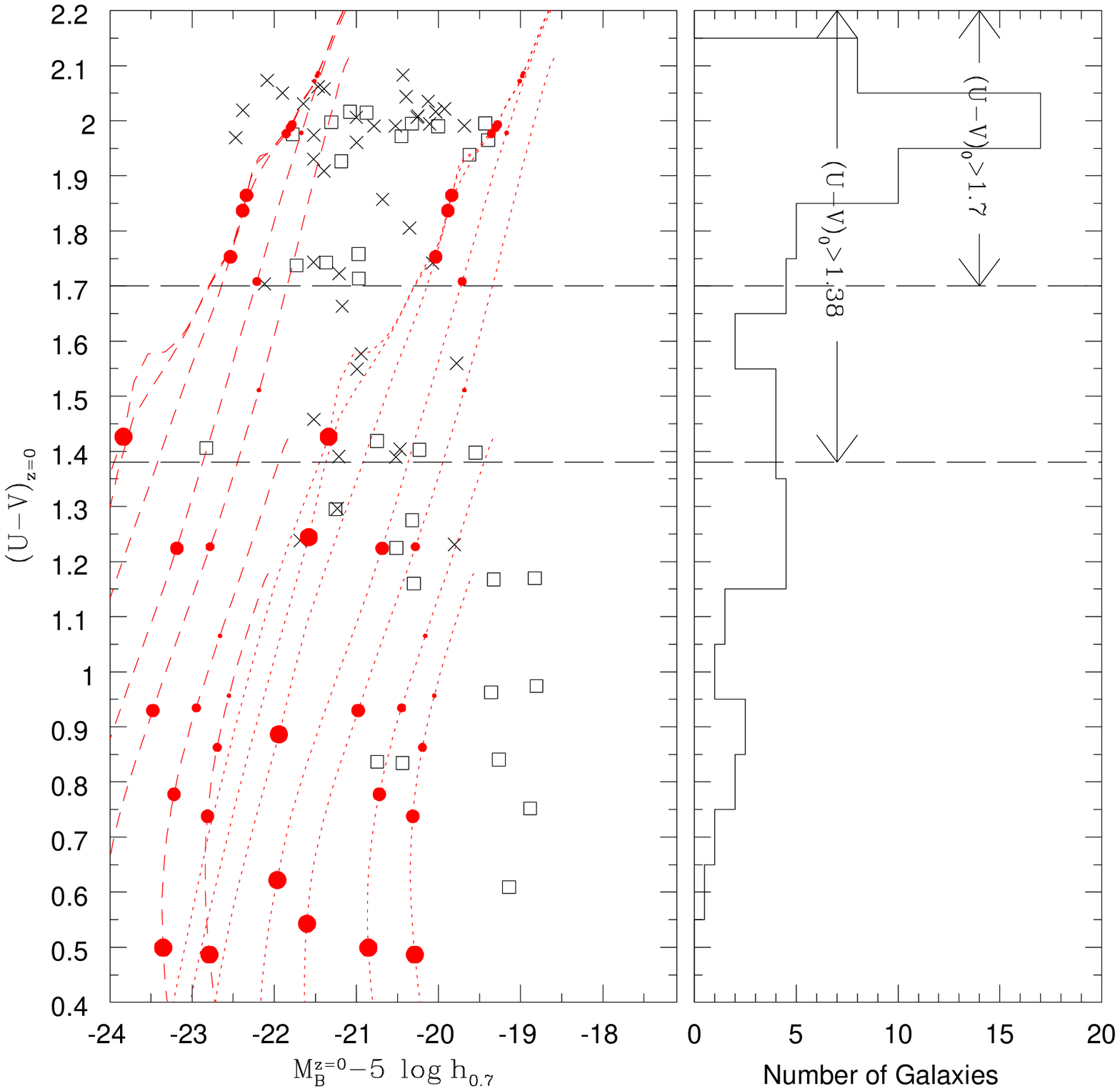,width=6.5cm}  
\epsfig{figure=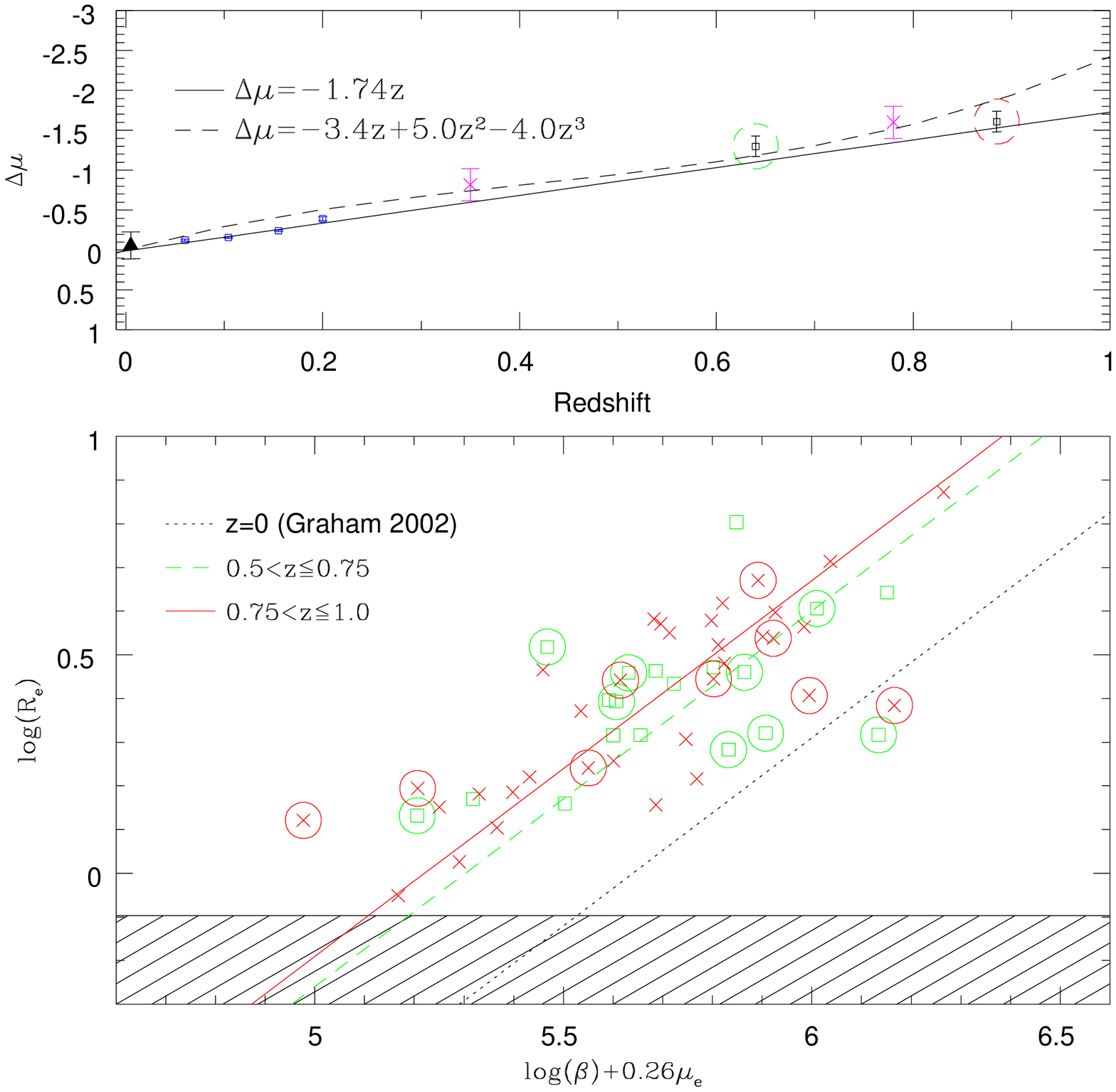,width=6.5cm}  
\end{center}
\caption{The left-hand plot shows the colour-magnitude diagram (CMD) and the 
right hand plot shows the photometric-plane diagram (PP) for
early-type galaxies. The CMD shows the distribution of rest-frame 
$(U-V)_0$ colour against $M_B^{z=0}$ for early-type galaxies with 
$0.5<z\le0.75$ (squares) and $0.75<z\le1.0$
(crosses).The red lines show the Bruzual \& Charlot \cite{BC03} 
evolutionary tracks for a $10^{11}{\cal M_{\odot}}$ (dotted) and
$10^{12}{\cal M_{\odot}}$ (dashed) galaxy with an exponentially
decaying star-formation rate with time-scales $\tau=0.1 - 9.0$ Gyr, 
from left to right, see \cite{Cr04}. The red circles represent the age of the galaxy increasing from 1 Gyr (largest circle) to 7 Gyr (smallest circle). 
The black dashed lines represent the selection criteria used to mimic
different color selections employed in the literature, see text.  
The right-hand panel shows the histogram in colour for the combined
sample. The lower panel of the PP shows our data with points as
above. Our best fit lines are the solid one for
$0.75<z\le1.0$ and the long-dashed for $0.5<z\le0.75$. The $z=0$ fit
from \cite{Grm02} is shown by the short 
dashed line. The blue E/S0s are marked by circles.  
In the top panel, we show the variation in surface-brightness with
redshift. Our points are marked by the squares ringed by 
large circles. The Schade \cite{Scd99} results are marked by magenta-crosses,
the SDSS \cite{Bnd03} results are marked by blue-squares and the
$z=0$ result is marked with the black-triangle. 
The solid line shows our best fit to these results.
\label{fig:Mcol}
} 
\end{figure} 

\vspace{-5mm} 
\section{Data}

 For details on all the fields, methods and equations, see the 
published paper \cite{Cr04}.
The data were extracted from 5 fields observed by the ACS as part of
GTO observations. The fields were selected to give 
accurate photometric redshifts (3 or more filters), to not have any primary 
targets in the range $0.5<z<1.0$ and to not contain any strong lensing
clusters at lower redshifts. The 5 targets were NGC 4676, UGC 10214,
TN J1338 $-$1942, TN J0924 $-$2201 and the Hubble Deep Field North
(HDFN). The combined area of these fields is $47.9\arcmin^2$. 

Our final sample contains 72 galaxies, 10 of which have 
spectroscopic redshifts, and the rest have photometric redshifts
determined using BPZ \cite{BPZ}. We calculate the rest-frame 
Johnson B-band from the ACS $i_{775}$ or $I_{814}$ bands, where the
k-correction is minimal at $z\sim0.75$. Both bands are used
depending on the dataset. Converting from these filters to the 
$z=0$ Johnson B-band removes any differences particular to the
passband. 

We calculate the half-light radius $r_e$ and total magnitude of each
galaxy using GALFIT \cite{GFIT}. In each case we fitted a single
Sersic profile. An initial sample was morphologically classified using a
semi-automated method, combining an eyeball classification to remove
galaxies that are obviously not E/SO and an automatic selection of those with
Sersic parameter $\beta>2.0$ and $r_e>0.1\arcsec$. We find that a
limit of $B<24.5$ for $z<0.75$ and $B<24.0$ for
$z>0.75$, $I\sim24$, gives us reliable morphologies and photometric
redshifts. A redshift range $0.5<z<1.0$ is limited by reliable
k-corrections and photometric redshifts.

The $0.5\le z<0.75$ sample contains 32 galaxies and the $0.75\le z<1.0$ 
sample contains 40 galaxies. Since our samples are morphologically selected 
rather than colour or SED selected we will be able to study the colour 
evolution of the galaxies. 

\vspace{-5mm} 
\section{Properties of Early Type Galaxies}

An unbiased look at the colours of E/S0 galaxies is important, not only
for understanding their star formation history, but also for
understanding the role that colour selection has in isolating large
samples of these objects at high redshift. Such colour (or SED)
selections have already been employed in the CFRS, CADIS, and COMBO-17
surveys and are relatively cheap to perform, requiring only
ground-based imaging over large areas of the sky.  Morphologies and
structural properties are, by contrast, much more expensive to
acquire, requiring the unique high resolution capabilities of HST.

The left-hand plot of Fig.~\ref{fig:Mcol} shows the absolute 
B-band magnitude against the rest-frame $(U-V)_{z=0}$ (AB) colour. 
We find a significant range in colours of early-type galaxies, with the 
majority having $(U-V)_0>1.7$. Those with $(U-V)_0>1.9$ have colours similar 
to the classic red ellipticals \cite{Bll04}. 
The red colours are consistent with an old 
coeval population of stars. While there is a slight colour-magnitude 
relationship for $(U-V)_0>1.9$ galaxies, the red sequence is blurred by a 
combination of the wide redshift range and errors in the photometric 
redshifts. For the remainder of the paper we define galaxies with 
$(U-V)_0>1.7$ as `red' and galaxies with $(U-V)_0<1.7$ as `blue'.

There are a large number of blue early-type galaxies. These have a 
broad colour distribution, implying a wide range in age or 
metallicity, with some ongoing star-formation. There is also a wide range
in absolute magnitude for $(U-V)_0>1.2$, $-22.5<M_B<-18$, but only the
very faintest galaxies have $(U-V)_0<1.2$. 

Fig.~\ref{fig:Mcol} also shows the expected evolutionary tracks of galaxies 
with different masses and decay timescales. Galaxies undergoing pure 
luminosity evolution with an exponentially decaying star-formation rate 
as described above will move along these tracks from blue to red. The tracks 
show that these galaxies, regardless of the decay timescale, reach a maximum 
B-band luminosity at $(U-V)_0<0.7$ and then they gradually fade as they 
redden.  While most of the $(U-V)_0>1.7$ 
E/S0s have $M>10^{11}{\cal M_{\odot}}$ and some have $M>10^{12}{\cal 
M_{\odot}}$, the bluer E/S0s, $1.2<(U-V)_0<1.7$, have $10^{10}<M<10^{11}{\cal 
M_{\odot}}$ and those with $(U-V)_0<1.2$ have only $M<10^{10}{\cal 
M_{\odot}}$. Note that these results should be treated with caution 
given their obvious dependence on our simple exponentially decaying
model. The very brightest of the blue E/S0s will end up amongst the 
red sequence that has already formed, but most will end up extending
the sequence to fainter absolute magnitudes, given pure luminosity
evolution. Dry merging amongst elliptical galaxies may also help produce
more luminous red ellipticals in the local universe \cite{Fbr05}.

Early-type galaxies have gone through a period of high star-formation, 
and the youngest of these galaxies are being systematically 
missed by ground-based surveys that select by
colour or SED, rather than morphology \cite{Cr04}. 

When we look at the structural properties of galaxies it is important to 
understand the selection effects \cite{Cr04}. To compare galaxies
within each redshift range, we use 
a sample that is volume-limited from $0.5<z\le1.0$, with $M_B\le-20.1$ mag 
and $R_e>0.8$ kpc.  

To study the structural parameters, we look at the distributions of
the half-light radius $R_e$ and the Sersic parameter $\beta$. To
compare each distribution we calculate the biweight and
biweight-scale. In both cases there is
no change between the two 
redshift ranges for the biweight size $<R_e>=2.6\pm0.2$ kpc, and the
biweight Sersic parameter $<\beta>=4.4\pm0.4$ and there is no 
significant difference in the biweight sizes of red or blue early-types in 
either redshift range. However, blue galaxies have a steeper Sersic
profile than red galaxies ($<\beta>=4.7$ vs $<\beta>=4.2$). The larger 
values of $\beta$ are consistent with bluer galaxies having a
starburst in the cores: the central regions will be slightly brighter, 
making the galaxies appear more concentrated \cite{Mnt99}. However, at 
fainter luminosities $-20.1<M_B<-18.8$, the distributions become 
significantly different, particularly in the Sersic parameter. Red
E/S0s at lower luminosities are slightly smaller, with $<R_e>\sim2.0$
and $<\beta>\sim4.1$. Blue E/S0s at lower luminosities have $<R_e>\sim2.0$
and $<\beta>\sim2.7$.
 
We find that our data has a good fit to the photometric plane
\cite{Grm02}. The photometric plane for our data is plotted in the 
right hand side of Fig.~\ref{fig:Mcol} and compared to the $z=0$
result \cite{Grm02} result. 

There is significant change in the offset of the photometric plane
with redshift. There is also a small change in offset between the 
red and blue galaxies, which is due to the variation in the Sersic 
parameter with colour. Since we have already shown that both $R_e$ and
$\beta$ have no significant evolution, the evolution in the
photometric plane must be due to evolution in $\mu_e$. In the top
panel of the figure, we calculate the change in $\mu_e$ compared to $z=0$
\cite{Grm02}. We also plot results from other surveys, the Sloan
Digital Sky Survey \cite{Bnd03} and the DGSS \cite{Scd99}, 
at $z=0.35$ and $z=0.78$. The variation is linear with redshift, 
$\Delta\mu=-1.74z$. Our results are similar the fundamental plane
results for E/S0 galaxies in the DGSS, \cite{Gbh03}.

\vspace{-5mm} 
\section{The Luminosity Function of E/S0 Galaxies}

We calculate luminosity functions for 3 different samples in both of
the redshift ranges \cite{Cr04}. The first sample is
our full sample, which is a morphological selection. Our second sample
is a 'red' sample, with $(U-V)>1.38$ (in addition to the morphological
selection) and our final sample is a 'very red' sample, with
$(U-V)>1.7$. We fit Schechter functions to the $0.5<z<0.75$ LFs and
use the faint-end slope value ($\alpha$) to constraint the
$0.75<z<1.0$ LFs. 

\begin{figure}  
\vspace*{1.25cm}  
\begin{center}
\epsfig{figure=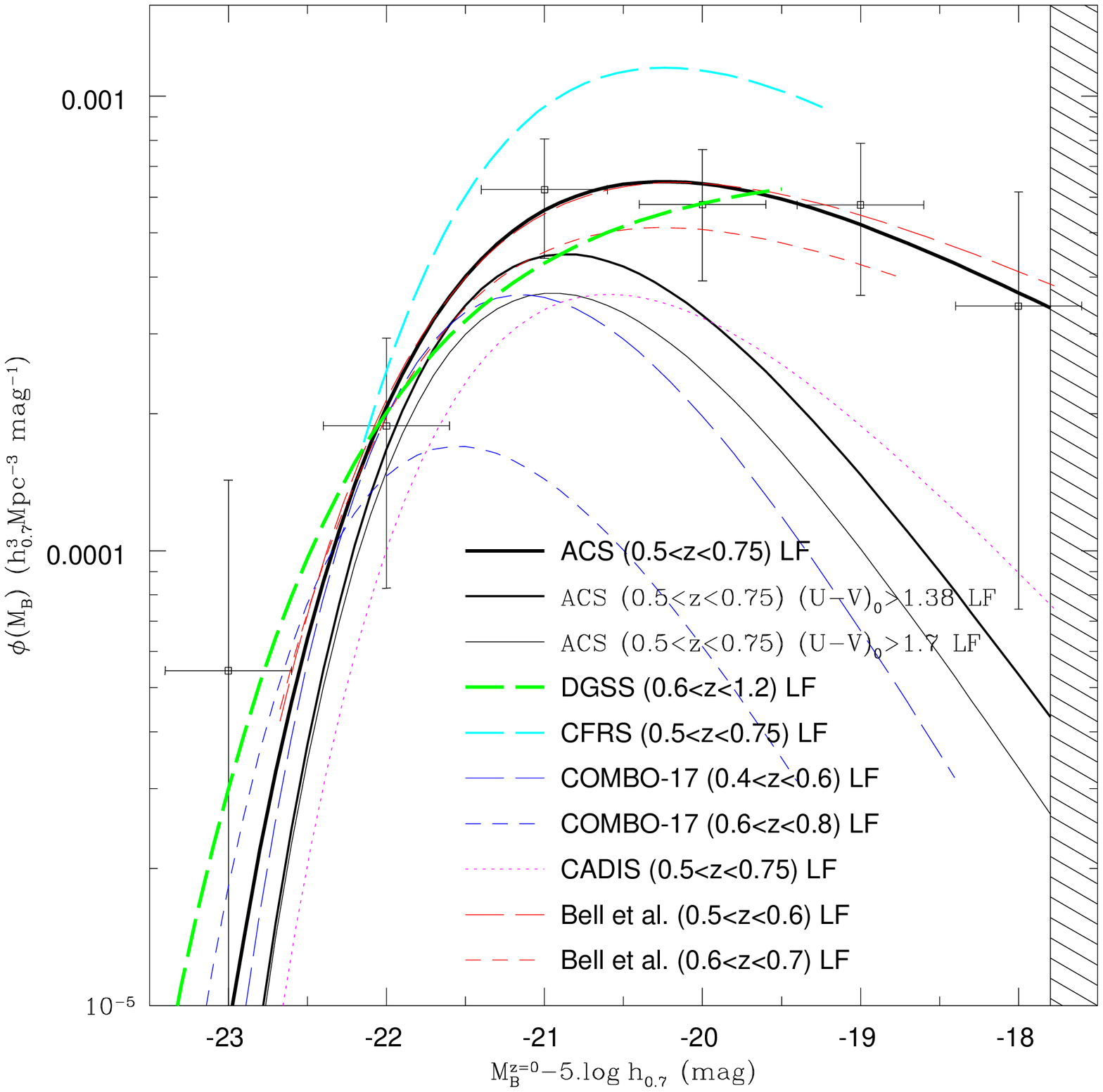,width=6.5cm}  
\epsfig{figure=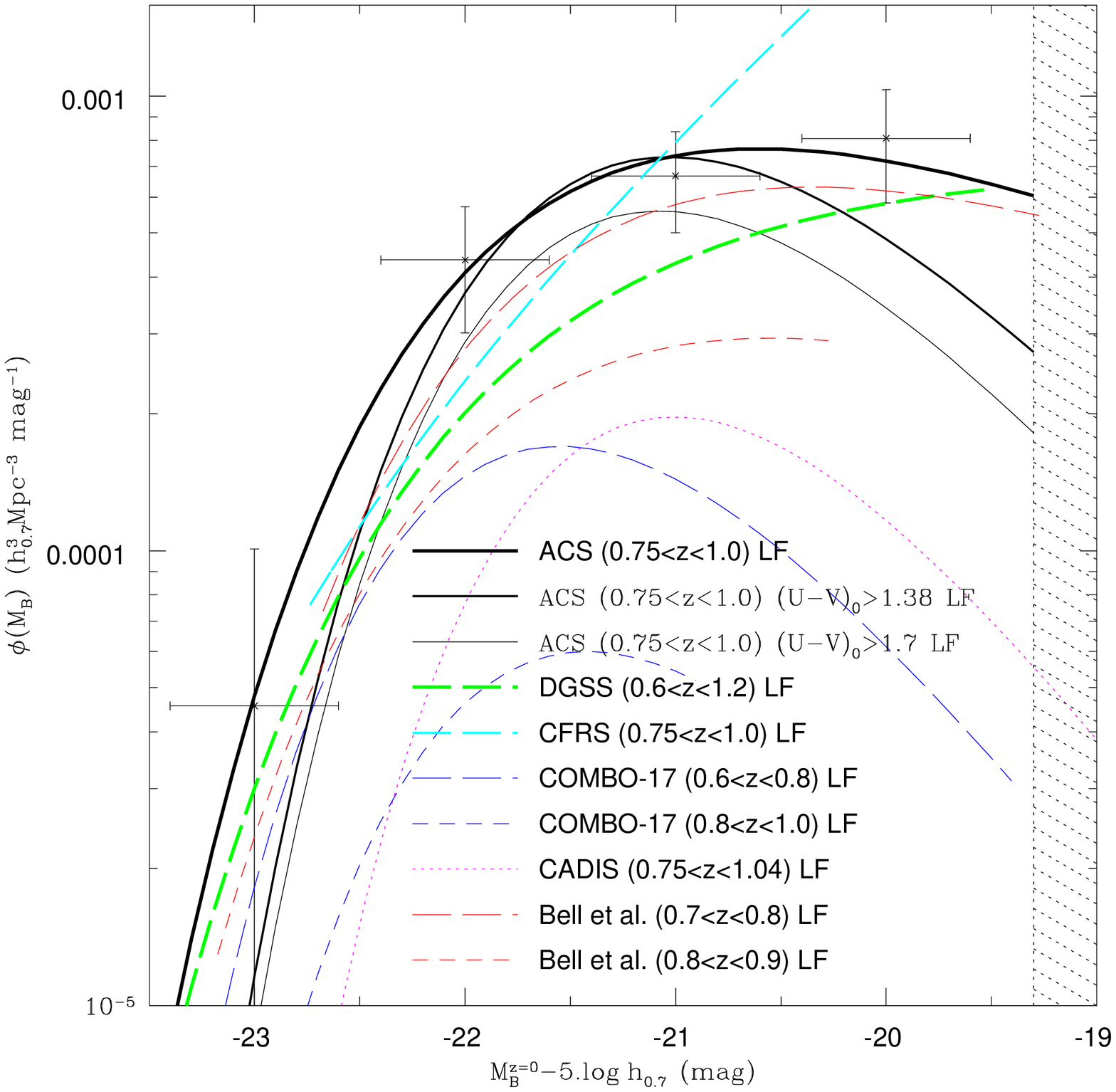,width=6.5cm}  
\end{center}
\vspace*{0.25cm}  
\caption{The luminosity functions of our $0.5<z<0.75$ (left) and
$0.75<z<1.0$ (right)
early-types compared to that from previous surveys. The ACS LFs are
plotted with black solid lines as indicated. The LFs for various
surveys are shown by the Schechter function fits as indicated.
All the luminosity functions have been converted to a $\Lambda$-CDM cosmology
with $H_0=70$ km s$^{-1}$ Mpc$^{-1}$.\label{fig:LF} 
} 
\end{figure}

In Fig.~\ref{fig:LF} we compare the LFs with each other and 
with other rest-frame B luminosity functions for early-type
galaxies. They are split into two redshift ranges, $0.5<z<0.75$ (left) 
and $0.75<z<1.0$ (right). In each case, we have converted from the
given cosmology to 
$\Omega_M=0.3$, $\Lambda=0.7$, $H_0=70\,$km s$^{-1}$Mpc$^{-1}$ and the
magnitudes to AB magnitudes. The points with error-bars are 
those for our morphologically selected luminosity function. The 
black lines show our LFs, with the thick line showing the morphologically
selected sample, the medium-thick line showing the 
$(U-V)_0>1.38$ sample and the thin line showing the $(U-V)_0>1.7$ sample.
The main difference is between the morphologically-selected sample and the 
colour-selected sample, with little difference between the
$(U-V)_0>1.38$ sample and the $(U-V)_0>1.7$ sample. This difference
occurs at the faint end, where most of the
very blue galaxies are. The blue lines 
with long dashes and short dashes show the original COMBO-17 LFs
\cite{Wlf03} at $z=0.5$ and $z=0.7$, respectively ($z=0.7$ and
$z=0.9$ respectively in the right hand plot), while the
equivalent red lines show the COMBO-17 LFs with new
colour-selections \cite{Bll04}. The magenta-line shows
the CADIS LF \cite{Frd01}. The cyan line represents the CFRS
\cite{Lll95} LF, and the green line shows the DGSS \cite{Im02}
LF. The original COMBO-17 and CADIS LFs select objects classified as
E-Sa from SED templates
and should be best matched to the $(U-V)_0>1.7$ sample. 
The CFRS should be best matched to the $(U-V)_0>1.38$ sample and the 
DGSS is morphologically selected and so can be compared to the full
sample. Overall we find that our LFs evolve with a  $0.36\pm0.36$ mag 
decrease in luminosity and ($15\pm12\%$) decrease in number density
from $0.75<z<1.0$ to $0.5<z<0.75$. The faint-end slope becomes
shallower as the colour-selection becomes stricter, increasing from
$\alpha=-0.53$ to  $\alpha=0.24$ to $\alpha=0.35$.
  
The present study has similar luminosity functions to the DGSS, see 
Fig.~\ref{fig:LF}, especially in the lower redshift
sample. The DGSS was not able to constrain the 
faint end slope, so they used a value $\alpha=-1.0$, based on the 
morphologically selected low redshift luminosity functions \cite{Im02}. 
Our sample goes almost 2 magnitudes deeper than the DGSS and hence we 
are able to constrain the faint end slope: $\alpha=-0.53\pm0.17$ is 
shallower than the DGSS LF, but is much steeper than the 
colour-selected luminosity functions. 

When we compare the samples selected with $(U-V)_0>1.38$, we find that the
CFRS is not a good match to the ACS LF: the space density is about
twice as high in the
CFRS as our measurement. In the higher redshift range, the CFRS luminosity
function is a closer match for $M_B<-21$ mag, but again overestimates the
number of galaxies for lower luminosities. This suggests that there is some 
contamination by late-type galaxies such as 
Sa/Sbc spirals, which are removed by our morphological selection, as well as 
incompleteness to bluer early-type galaxies.

The $0.5<z\le0.75$ CADIS LF \cite{Frd01} and original $0.4<z\le0.6$ 
COMBO-17 LF \cite{Wlf03} both closely
resemble the ACS $0.5<z\le0.75$, $(U-V)_0>1.7$ LF, with offsets of
$\sim0.25$ magnitudes either way, which is within the errors. 
However the original $0.6<z\le0.8$ COMBO-17 LF has a much lower space density, 
which is also much lower than the ACS $0.75<z\le1.0$, $(U-V)_0>1.7$
LF. At higher redshift there are increasing disparities (by factors of
10) between the original COMBO-17 and the ACS LFs. The new COMBO-17
LFs \cite{Bll04}, which use a colour-selection
based upon the evolution of the red-sequence, give results that are
very close to our morphologically selected sample for their $0.4<z\le0.6$
and $0.6<z\le0.8$ LFs, but still underestimate the space density
at $0.8<z\le1.0$. 

Simple colour cuts, will inevitably lead to contamination from
spirals or miss many red ellipticals if passive evolution is not
taken into account. However, even when evolution is accounted for, the
blue ellipticals will be missed in a colour-selected sample. Bell
\cite{Bll04} has determined that they miss $20-30\%$ of the elliptical
population in their sample but also have about $20-30\%$
contamination in their $z=0.5$ sample. The proportions change at
higher redshift making the LFs more disparate. 

These results show that there is a wide variation in the luminosity
functions reported and that selection effects have a systematic effect
on the results.  In particular, for colour-selected samples, we
noted a significant underestimate of the faint end slope compared with
morphologically selected samples. The space density of $M^*$ galaxies
also varied greatly from survey to survey.

\vspace{-5mm} 
\section{Conclusions}

From the analysis of the colours and structural properties of E/S0 
galaxies at $0.5<z<1.0$, it is apparent that bright ($M_B<-20.1$), 
`blue' $(U-V)_0<1.7$ E/S0 galaxies are not significantly different from
bright `red' $(U-V)_0>1.7$ E/S0 galaxies in terms of their structural 
parameters, and there is no significant evolution in the structural
parameters. However, there is evolution in the luminosity ($\sim0.4$
mag) of these galaxies as demonstrated by the photometric plane and LFs.
At $z=0$, the blue E/S0s will be 
only slightly less luminous than the `red' galaxies, and there
will be significant overlap, see Fig.~\ref{fig:Mcol}. Fainter
($M_B>-20.1$) `blue' E/S0 galaxies are smaller 
with lower Sersic parameters than their `red' counterparts. These galaxies 
often have extremely blue colours $(U-V)_0<1.2$ and are likely to be
much less massive. The evolution tracks Fig.~\ref{fig:Mcol} 
suggests that these will fade by $\sim3$ mag as their stellar 
populations age. This is consistent with these galaxies becoming 
present day dwarf ellipticals. 

Using deep high resolution optical data we are able to measure the 
morphological E/S0 luminosity function almost 2 magnitudes deeper than the 
DGSS and constrain the faint end slope of the $0.5<z\le0.75$  LF. We find a 
fairly flat faint end slope $\alpha=-0.53\pm0.13$, slightly shallower than 
low redshift luminosity functions for morphologically selected E/S0s but
much steeper than colour-selected samples. 

Using purely photometric information (colour, SED) to select the galaxy 
sample misses the bluer early types, and may lead to 
contamination from Sa/Sbc spiral galaxies or other red galaxies. As shown 
in Fig~\ref{fig:LF} this leads to a large variation in the measurement of
the luminosity function which can lead to significantly different
conclusions on the evolution of galaxies.

\vspace{-5mm}  
\acknowledgements{ 
ACS was developed under NASA contract NAS5-32865 and this research has 
been supported by NASA grant NAG5-7697. The STScI is operated by AURA Inc., 
under NASA contract NAS5-26555. We are grateful to Ken Anderson, Jon 
McCann, Sharon Busching, Alex Framarini, Sharon Barkhouser, and Terry Allen 
for their invaluable contributions to the ACS project at JHU.
}

\vfill 
\end{document}